# Plasmon enhanced upconversion luminescence near gold nanoparticles – Simulation and analysis of the interactions

Stefan Fischer,[1,*] Florian Hallermann,[2] Toni Eichelkraut,[3] Gero von Plessen,[2] Karl W. Krämer,[4] Daniel Biner,[4] Heiko Steinkemper,[1] Martin Hermle,[1] and Jan C. Goldschmidt[1]

[1] *Fraunhofer Institute for Solar Energy Systems, Heidenhofstr. 2, 79110 Freiburg, Germany*
[2] *Institute of Physics (1A), RWTH Aachen University, 52056 Aachen, Germany*
[3] *Institute of Condensed Matter Theory and Solid State Optics, Abbe Center of Photonics, Friedrich-Schiller-Universität, 07743 Jena, Germany*
[4] *Department of Chemistry and Biochemistry, University of Bern, Freiestrasse 3, 3012 Bern, Switzerland*
[*]*stefan.fischer@ise.fraunhofer.de*

**Abstract:** We investigate plasmon resonances in gold nanoparticles to enhance the quantum yield of upconverting materials. For this purpose, we use a rate equation model that describes the upconversion of trivalent erbium based upconverters. Changes of the optical field acting on the upconverter and the changes to the transition probabilities of the upconverter in the proximity of a gold nanoparticle are calculated using Mie theory and exact electrodynamic theory respectively. With this data, the influence on the luminescence of the upconverter is determined using the rate equation model. The results show that upconversion luminescence can be increased in the proximity of a spherical gold nanoparticle due to the change in the optical field and the modification of the transition rates.

## 1. Introduction

The biggest losses of single band gap solar cells result from incomplete utilization of the broad solar spectrum. About 20% of the incoming solar energy is not utilized in silicon solar cells because the photons with energies below the band gap of silicon do not carry enough energy to generate free charge carriers. These photons are simply transmitted through the silicon. Upconversion (UC) of these low energy photons is a promising way to enhance the efficiency of solar cells [1, 2]. Analyses show that the theoretical upper efficiency limit of a silicon solar cell is raised from near 30% [3] up to 40.2% [4] for a silicon solar cell with an upconverter illuminated by non-concentrated light.

Hexagonal sodium yttrium fluoride ($\beta$-NaYF$_4$) doped with trivalent erbium (Er$^{3+}$), especially with a doping ratio of one erbium ion to four yttrium ions ($\beta$-NaEr$_{0.2}$Y$_{0.8}$F$_4$), is known for its very high quantum yield for UC of near infrared (NIR) photons at wavelengths around 1523 nm [5, 6]. An UC quantum yield of 4.3% at an irradiance of 1370 W m$^{-2}$ was determined by photoluminescence measurements [7].

UC silicon solar cell devices have been produced and investigated extensively. Mostly, lasers with rather high powers in the NIR have been used to measure a short-circuit current of the solar cell caused by upconverted photons [4, 6-9]. Recently, Goldschmidt et al. [10] observed comparable UC quantum efficiency for an upconverter solar cell device under broad band illumination with a xenon lamp, in comparison to monochromatic laser illumination with the same irradiance in the relevant spectral region. In all these experiments, however, the UC quantum yield remains too small to make UC relevant for photovoltaic applications. Therefore, additional means of increasing the UC efficiency are necessary. One promising possibility is to use the plasmon resonance in metal nanoparticles (MNP) in the proximity of the upconverting material.

For a long time, plasmonic effects have been intensively investigated to enhance luminescence, absorption and other properties of optical systems, such as Er$^{3+}$ [11-14]. The higher local field intensity caused by the plasmon resonance positively influences the UC efficiency because of the non-linear nature of UC. Additionally, the MNP also influence the transition probabilities of the luminescent system if transitions are resonant with the plasmon resonances [15-18]. This effect depends on the distance between emitter and metal nanoparticle. Enhanced decay rates by MNP are determined, for example, by lifetime measurements of $Eu^{3+}$ complexes on the surface of gold nanoparticles (GNP) that have been coated with silica ($SiO_2$) layers of different thicknesses to control the distance [19]. Lanthanide doped nanocrystals with Au nanoparticles on the surface have recently received a lot of attention because of the opportunity to modify the emission spectra by systematic amplification of certain transitions by the plasmon resonances [20]. Further, with the use of Au nanoparticles, the UC luminescence of $\beta$-NaYF$_4$:Yb,Tm nanocrystals could be enhanced by a factor of up to 118, depending on the transition and the incident irradiance [21]. An UC luminescence enhancement by up to a factor of 4.8 was determined for single $\beta$-NaYF$_4$:Yb,Er nanocrystals close to spherical GNP with a diameter of 60 nm [22].

In this work, we will analyze the effect of both local field enhancement and coupling of metal nanoparticle to the $Er^{3+}$, in order to change the transition probabilities. For this purpose, we developed a rate equation model to describe the UC. With experimentally determined parameters of $\beta$-NaEr$_{0.2}$Y$_{0.8}$F$_4$, a very good agreement between the simulations and the corresponding experiments was found [23]. The optical field enhancement of different spherical GNP was calculated in order to identify which diameter is particularly suitable for UC of 1523 nm photons. A spherical gold nanoparticle with a diameter of 200 nm showed sufficient resonance to the excitation wavelength of 1.5 µm and constitutes the reference model investigated in this work. Results of simulations for this spherical gold nanoparticle will be connected to the rate equation UC model. We will show how the optical field enhancement and the modification of the transition probabilities are introduced into the UC model and how these plasmonic effects influence the luminescence of the upconverter.

## 2. Upconversion model

The UC model considers ground state absorption (GSA), excited state absorption (ESA), stimulated emission (STE), spontaneous emission (SPE), energy transfer (ET) and multi-phonon relaxation (MPR). By ET we refer to energy transfer upconversion (ETU) shown in Fig. 1 (a) and cross-relaxation (CR), which is the inverse process. For CR, the arrows of ETU in Fig. 1 (a) are reversed. In the model, the occupation vector $\vec{n}$ and its rate of change $\dot{\vec{n}}$ are described by the following differential equation:

$$\dot{\vec{n}} = (M_{GSA} + M_{ESA} + M_{STE} + M_{SPE} + M_{MPR})\vec{n} + \vec{v}_{ET}(\vec{n},d) \qquad (1)$$

In this equation, $M_{GSA}$, $M_{ESA}$, $M_{STE}$, $M_{SPE}$, and $M_{MPR}$ are matrices that describe the different processes, while $\vec{v}_{ET}(\vec{n},d)$ is a vector describing ET. The different variables will be briefly explained in the following. A complete treatment of the rate equation model can be found in the work of Fischer et al. [23]. In Fig. 1 (a) the various processes are illustrated. Cooperative processes are not considered because they contribute only marginally [24].

The probability $W_{if}^{SPE}$ for SPE is given directly by

$$W_{if}^{SPE} = A_{if} \qquad (2)$$

where $A_{if}$ is the Einstein coefficient for SPE from an initial energy level $i$ to a final energy level $f$. The stimulated processes are connected to the $A_{if}$. The probability for GSA and ESA $W_{if}^{GSA/ESA}$ is calculated by

$$W_{if}^{GSA/ESA} = u(\omega_{if})B_{if} = u(\omega_{if})\frac{\pi^2 c^3}{\hbar \omega_{if}^3}\frac{g_f}{g_i}A_{fi} \qquad (3)$$

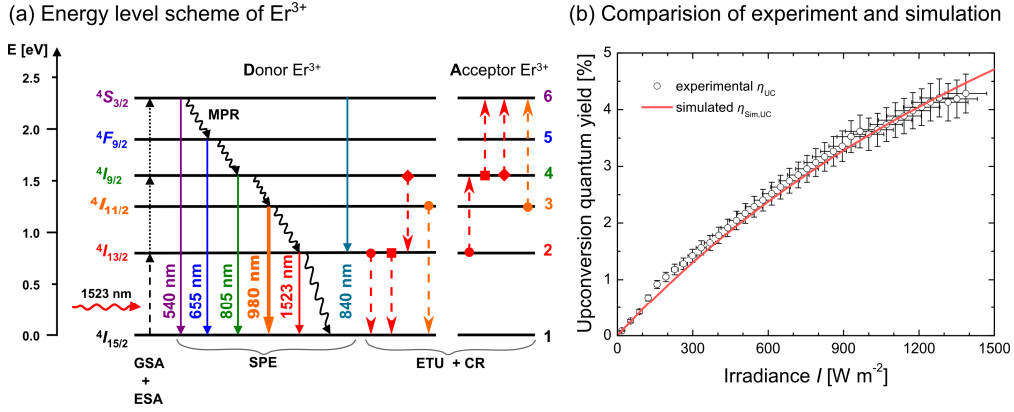

Fig. 1. (a) Energy levels of the trivalent erbium in the host crystal NaYF$_4$ with corresponding luminescence wavelengths of transitions from the excited states to the ground state (colored solid arrows). In the rate equation model, we assume an excitation with a wavelength of 1523 nm (red waved arrow). Higher energy levels can be populated by GSA followed by ESA (black dashed arrows). Another possibility to populate higher energy levels is ETU (dashed colored arrows), where energy is transferred from one ion, the donor (D), to a neighboring second ion, the acceptor (A). Cross-relaxation (CR) is the inverse process and included in the model as well. Additionally, we consider MPR to next lower energy levels (waved arrows). (b) Comparison of the simulated UC quantum yield $\eta_{Sim,UC}$ (red solid line) with the UC quantum yield, which was determined from calibrated photoluminescence measurements (black squares) [7]. Due to the non-linearity of UC, the UC quantum yield increases with increasing irradiance. Details on the UC model can be found in [23].

where $u(\omega_{if})$ is the spectral photon energy density, $B_{if}$ is the Einstein coefficient for stimulated processes from an initial energy level $i$ to a final energy level $f$, $\omega_{if}$ is the angular frequency of the transition, $g_f$ and $g_i$ are the degeneracies of the energy levels $i$ and $f$, and $c$ is the speed of light in vacuum [25]. The probability for STE $W_{if}^{STE}$ is

$$W_{if}^{STE} = \frac{\pi^2 c^3}{\hbar \omega_{if}^3} u(\omega_{if}) A_{if} = u(\omega_{if}) B_{if}. \tag{4}$$

In our experiments and for the presented simulations, the upconverter is excited by monochromatic, incident radiation at 1523 nm with an irradiance of 1000 Wm$^{-2}$. Therefore, the model was designed for spectral irradiance $I_v(\omega)$ instead of the spectral energy density $u(\omega_{if})$. In the frequency interval $d\omega$, these two quantities are related to each other via

$$I_v(\omega) d\omega = \frac{c}{n} u(\omega) d\omega \tag{5}$$

with $n$ being the refractive index of the upconverter.

The determination of the simulation parameters constitutes the main challenge for the generation of a quantitative model. The absorption coefficient $\alpha(\lambda)$ is, in principle, a quantity that is easy to determine. Since the upconverter sample $\beta$-NaEr$_{0.2}$Y$_{0.8}$F$_4$ is a microcrystalline powder, strong scattering occurs. This makes a direct optical measurement challenging, as the average optical path length does not correspond to the geometrical sample thickness and is not known precisely. Therefore, we used the revised Kubelka-Munk theory [26, 27] to determine $\alpha(\lambda)$ of our samples from diffuse reflection measurements of samples of different thicknesses [23].

If a lanthanide ion is at a site in a low-symmetry environment, a mixing of opposite parity states occurs and the selection rules, particularly for electric dipole transitions, are relaxed. This effect was investigated, independently, by Judd and Ofelt [28, 29]. Further, the Judd-Ofelt theory relates the absorption coefficient to the electric dipole matrix elements, which are correlated to the Einstein coefficients $A_{if}$ of the transitions that govern the dynamics of the UC processes. In consequence, with the absorption coefficient $\alpha(\lambda)$ all required transition probabilities for our rate equation model can be determined.

At this point, the model has only been discussed to an extent as to enable an understanding of how a plasmon resonance influences UC and of how the presented results are calculated. A more detailed description of the rate equation model and the experimental determination of the input parameters can be found elsewhere [23, 30].

*2.1. Results of the upconversion model*

For different irradiances, the occupation of the energy levels was simulated with the rate equation model. The luminescence $L_i$ of a certain energy level $i$ was calculated by multiplying the relative occupations $n_i$ by the corresponding Einstein coefficient for spontaneous emission:

$$L_i = n_i A_{if} \qquad (6)$$

The sum over the $L_i$ of all considered energy levels divided by the sum over the rate of all stimulated processes is proportional to the UC quantum yield $\eta_{UC}$

$$\eta_{\text{Sim,UC}} = \frac{\sum_{i=3}^{6} n_i A_{i1} + n_6 A_{62}}{\sum_{i \neq f} n_i M_{\text{GSA}} + n_i M_{\text{ESA}} - n_f M_{\text{STE}}} \propto \eta_{UC} \qquad (7)$$

and is called the simulated internal UC quantum yield $\eta_{\text{Sim,UC}}$. The sum over the indices $i \neq f$ of the matrices for stimulated processes is equivalent to the rate of absorbed photons minus the rate of the photons lost by STE at the excitation wavelength. The rate of the STE is marginal, because the occupations of higher energy levels are minor, hence only few $Er^{3+}$ could possibly contribute to STE. The $A_{i1}$ are the Einstein coefficients for spontaneous emission to the ground state and the sum from $i = 3$ to 6 indicates that only the energy levels with energy gaps above the band gap of silicon are considered. Therefore, the $\eta_{\text{Sim,UC}}$ can be compared to the experimentally determined UC quantum yield. In Fig. 1 (b) it can be seen that the simulation matches the shape of the experimentally determined relation between irradiance and UC quantum yield very well. Details on the experimental method can be found in Ref. [7]. The transition from $^4S_{3/2}$ to the first excited state $^4I_{13/2}$ at a corresponding wavelength of 840 nm is considered in this work as well, but the fractional contribution of this transition to the total luminescence is negligible. In both the simulation and the experiment, the transition from the energy level $^4I_{11/2}$ to the ground state $^4I_{15/2}$ contributes roughly 99% of the total UC luminescence, depending on the irradiance [7]. Therefore, the focus of the investigation on plasmon enhanced UC will be on this transition.

## 3. Plasmon enhanced upconversion

*3.1. Interactions of metal nanoparticle and upconverter within the rate equation model*

The simulations of the gold nanoparticle were performed independently from the UC model. Fig. 2 (a) shows a schematic of the simulation design. A spherical gold nanoparticle is located in the center of a cubic simulation volume with an edge length of 6 times the diameter of the nanoparticle. The incident light comes from the bottom of the cube; the same applies to the graphs shown later in this work. We present results for a spherical gold nanoparticle with a diameter of 200 nm, which results in a cube with an edge length of 1200 nm.

The UC model was coupled with simulations of plasmon resonances of spherical GNP. The optical field scattered by the metal nanoparticle was calculated using the Mie theory. The local change in the optical field affects the spectral photon energy density $u(\omega_{if})$. The change in the optical field thus modifies all stimulated processes; more precisely the GSA, ESA, and STE. The change in the square of the optical field was calculated in comparison to the case without the nanoparticle, in order to obtain a relative enhancement factor $\gamma_E$. This enhancement factor $\gamma_E$ was calculated for every position in the simulation volume. Fig. 2 (b) shows the relative enhancement for the x-z-plane at the position of 0 nm on the y-axis. The color scale in Fig. 2 (b) runs from 0 to 5 in order to also show the structure of the field enhancement further away from the particle. Very close to the particle field, enhancement factors as high as 16 are achieved. To calculate the effect due to the field enhancement, the probabilities of the stimulated processes calculated with equations (3) and (4) need to be multiplied by $\gamma_E$.

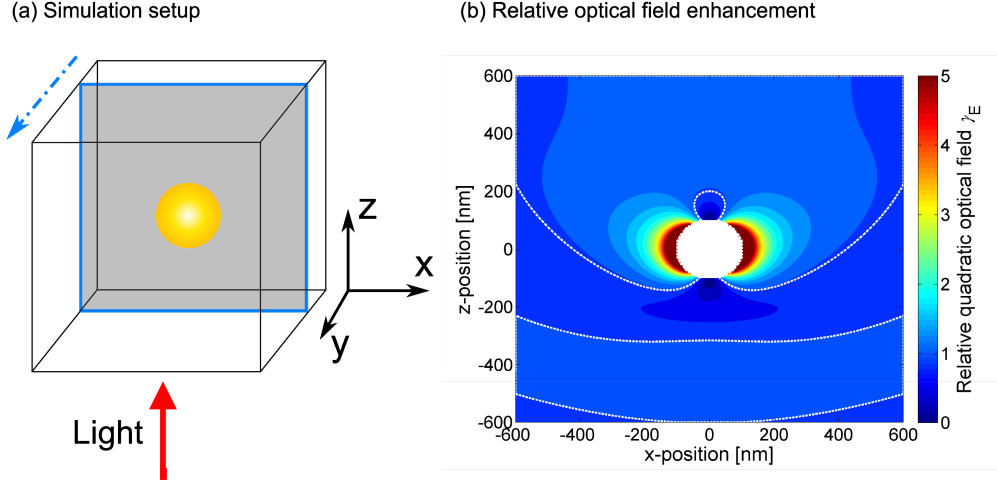

Fig. 2. (a) The simulation was performed in a cubic volume with an edge length of 6 times the diameter of the gold nanoparticle, which is located in the center of the cube. The relative luminescence intensities of the various transitions of the upconverter were calculated for every position in the simulation volume. The values in the x-z-plane were calculated for every value of the y-axis. For a clear presentation, the graphs in this work show the x-z-plane at y = 0 nm, the center of the cube. (b) Enhancement of the quadratic optical field $\gamma_E$ due to a spherical gold nanoparticle with a diameter of 200 nm (white circle in the center). The color scale was clipped at 5 for a clearer presentation and the white dashed line represents $\gamma_E = 1$. The strongest field enhancement with a value of 16 is reached close to the surface of the metal nanoparticle. The plot shows a cut in the x-z-plane of the cubic simulation volume in the center at y = 0 nm.

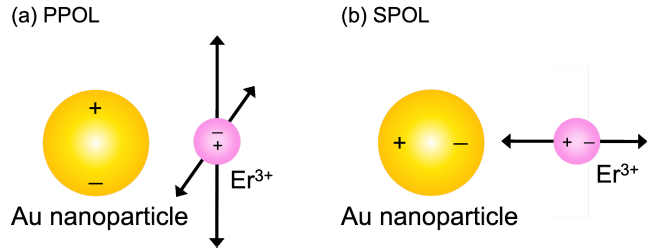

Fig. 3. Schematic of the two polarizations considered for the simulations. For PPOL the $Er^{3+}$ dipole oscillates parallel to the surface of the gold nanoparticle, while for SPOL the $Er^{3+}$ dipole oscillates perpendicularly to the surface of the gold nanoparticle.

The second effect of the gold nanoparticle is that it influences the transition probabilities within the $Er^{3+}$ upconverter. The changed density of states in the presence of a metal nanoparticle is responsible for the enhanced absorption and emission probabilities. In our model this is described by the modification of the Einstein coefficients $A_{if}$, which affects all processes except MPR. The relative change in the radiative transition probabilities $\gamma_{rad}$ and in additional non-radiative transition probabilities $\gamma_{nonrad}$ are calculated with exact electrodynamic theory in which the $Er^{3+}$ is treated as a dipole emitter [16, 31, 32]. The additional non-radiative transition probability considers energy transfer from the $Er^{3+}$ to the metal nanoparticle where the energy is dissipated. Details on the calculations can be found in [14, 32]. The modified Einstein coefficients $A_{if,\text{plasmon}}$ are

$$A_{if,\text{plasmon}} = (\gamma_{\text{rad}} + \gamma_{\text{nonrad}})A_{if} \tag{8}$$

and replace the former $A_{if}$ in the equations described above. To take care of the additional non-radiative losses, the luminescence is calculated by

$$L_{i,\text{plasmon}} = \gamma_{rad} n_i A_{if} . \tag{9}$$

The orientation, also referred to as polarization, of the $Er^{3+}$ dipole emitter with respect to the surface of the gold nanoparticle has a very strong influence on the coupling. We distinguish between two polarizations of the oscillating $Er^{3+}$ dipole: parallel (PPOL) and perpendicular (SPOL) to the surface of the gold nanoparticle. A schematic of the polarizations is shown in Fig. 3.

For both polarizations the $\gamma_{\text{rad}}$ and the $\gamma_{\text{nonrad}}$ need to be calculated. The coupling of nanoparticle and the $Er^{3+}$ only depends on the distance between them and not on other spatial orientations. In contrast, the $\gamma_E$ strongly depends on the spatial position of the $Er^{3+}$ with respect to the metal nanoparticle.

### 3.2. Results

The rate equations of the UC have been solved for partial cubic simulation volumes with an edge size of 5 nm with the modified transition probabilities ($\gamma_{\text{rad}}$, $\gamma_{\text{nonrad}}$) and the relative enhancement of the local quadratic optical field ($\gamma_E$). For the investigated upconverter, approximately 125 $Er^{3+}$ are in such a partial volume. This amount of $Er^{3+}$ is sufficient to solve the rate equations in the partial volumes reasonably. For much less $Er^{3+}$, diffusion effects of the excited electrons into neighboring partial volumes are much more likely and need to be considered. The luminescence of the energy levels was determined via equation (6) and normalized to the case without the gold nanoparticle. This relative luminescence, often called the enhancement factor, describes how strong the gold nanoparticle increases (>1) or decreases (<1) the emission from certain transitions.

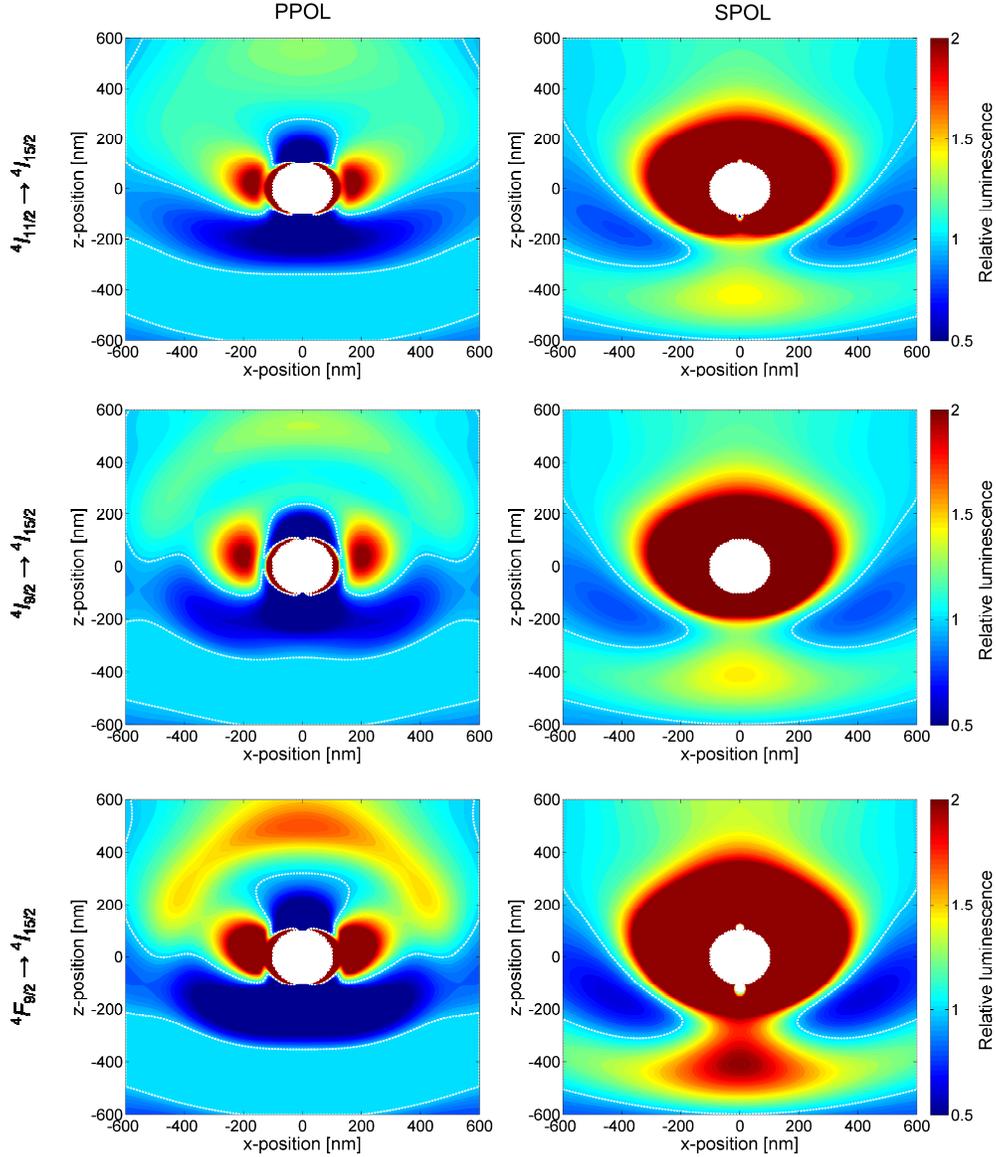

Fig. 4. Impact of a spherical gold nanoparticle with a diameter of 200 nm on the relative luminescence of certain transitions. The rate equations of the UC model were solved for every spatial point, taking into account the relative change in the quadratic optical field and the relative changes of the transition probabilities due to the coupling of metal nanoparticle and the $Er^{3+}$ dipole. The color scale shows the relative luminescence, which was calculated by dividing the luminescence with metal nanoparticle by that without the metal nanoparticle. The dashed white lines represent enhancement factors of 1. Two polarizations, PPOL and SPOL, were considered (see Fig. 3). The transitions from the energy levels $^4I_{11/2}$, $^4I_{9/2}$ and $^4F_{9/2}$ to the ground state $^4I_{15/2}$ in the x-z-plane at y = 0 nm are shown. The white circle in the middle of the graphs represents the spherical gold nanoparticle. For a clearer presentation, the color scale was clipped below 0.5 and above 2.

### 3.2.1. Spatial resolved relative upconversion luminescence

The relative luminescence from the transitions of the energy levels $^4I_{11/2}$, $^4I_{9/2}$ and $^4F_{9/2}$ to the ground state $^4I_{15/2}$ is shown in Fig. 4. On the left side are the graphs for PPOL and on the right side for SPOL. The results are shown for the x-z-plane at y = 0 nm. Since the field enhancement $\gamma_E$ affects the upconverter equally for both polarizations of the emission, the strong differences between PPOL and SPOL are attributed to the coupling of the metal nanoparticle and the $Er^{3+}$ dipole. With SPOL, very high enhancements can be reached near the nanoparticle and the structure of the relative luminescence for the different transitions is very similar. The strongest enhancement for the transition from $^4I_{11/2}$ to $^4I_{15/2}$ is localized at a certain position close to the gold particle, where the relative luminescence is 500 times higher than without the gold particle. For the transitions from $^4I_{9/2}$ and $^4F_{9/2}$ to the ground state, even higher enhancement factors can be reached, with values of 4440 and 2990, respectively.

For PPOL, the relative luminescence is much lower compared to the SPOL, but still higher than without the nanoparticle. Furthermore, for PPOL the structure of the relative luminescence is very different for the different transitions. There are shells around the gold nanoparticle at a distance of approximately 450 nm, where an increase in the relative luminescence is indicated. The highest enhancement factors reached are 9, 155 and 240 for the transitions from $^4I_{11/2}$, $^4I_{9/2}$ and $^4F_{9/2}$, respectively, to the ground state $^4I_{15/2}$.

*3.2.2. Averaged over a certain distance to the surface of the nanoparticle*

The presented results indicate that both a strong increase and a decrease in UC luminescence are caused by spherical GNP depending on the precise position in the investigated volume. Positioning of the UC at the best suited position is technologically very demanding. On the other hand, a coating of the MNP that ensures a certain distance is a feasible option. Therefore, we analyzed the dependence of the relative luminescence on the distance to the metal nanoparticle. At certain distances, the average values over spherical shells around the metal nanoparticle were calculated. In Fig. 5 (a) and Fig. 5 (b) the average relative luminescence in dependence of the distance to the metal nanoparticle are shown for PPOL and SPOL, respectively. For PPOL, the dominant UC luminescence from $^4I_{11/2}$ generally decreases, when averaged over many positions at the same distance to the metal nanoparticle. There is a peak at a distance of 15 nm, where the relative luminescence increases slightly to 1.04. Higher transitions benefit from the presence of the gold nanoparticle, and very high luminescence enhancements close to the nanoparticle can be reached.

On the other hand, for SPOL the averaged increase in the luminescence is very large. For the transition from $^4F_{9/2}$ to $^4I_{15/2}$, a factor of 490 is reached at a distance of 15 nm. The enhancement drops to 135 for a distance of 100 nm and further to 10 at 400 nm. However, for the dominant transition from $^4I_{11/2}$, an increase of more than 50 is determined at a distance from 10 nm to 50 nm.

The large difference between the two polarizations of the dipole emitter indicate the large impact of the changed transition probabilities due to the modification of the density of states. Additionally, there is an enhanced relative luminescence far away from the metal nanoparticle where the optical field is only marginally enhanced or even reduced. Since this effect is attributed to the changed transition probabilities as well, further optimization of the MNP should also focus on these modifications.

To derive an overall result, the average over SPOL and PPOL was calculated. Because there are two possible orientations for PPOL, the PPOL orientation was weighted with a factor of two in the average. The results are presented in Fig. 6. It can be seen that also for this average high enhancement factors, as high as 170 for the transition from $^4F_{9/2}$, are feasible. For the transition from $^4I_{11/2}$, the luminescence can be enhanced by a factor of more than 15 at distances between 10 nm and 40 nm. Even for larger distances a positive effect of the spherical gold nanoparticle on the UC is determined, which is attributed to the changed transition probabilities due to the altered density of states.

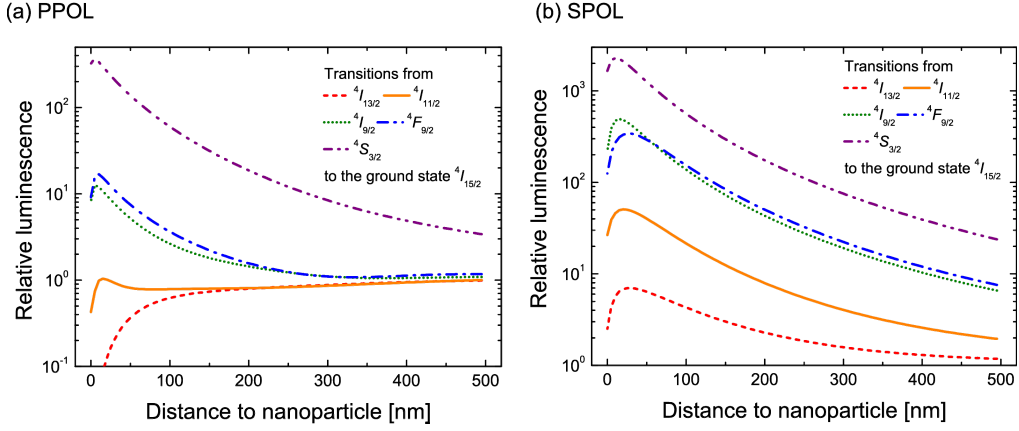

Fig. 5. (a) Relative luminescence of various transitions of the UC for PPOL in dependence of the distance to the surface of the gold nanoparticle. While the luminescence of the two lowest transitions from $^4I_{13/2}$ and $^4I_{11/2}$ decrease slightly, the higher transitions with lower wavelength benefit from the nanoparticle. (b) Relative luminescence of various transitions of the UC for SPOL in dependence of the distance to the surface of the gold nanoparticle. For all distances and all transitions, the GNP increase the luminescence. For the transition from $^4I_{11/2}$, the best suited distance is between 10 nm to 50 nm, where the luminescence is enhanced by a factor of 50. The strongest enhancement is reached for the transition from $^4F_{9/2}$ with a value of 2270 for a distance of 10 nm to the surface of the gold nanoparticle. However, this transition contributes only marginally to the overall UC luminescence.

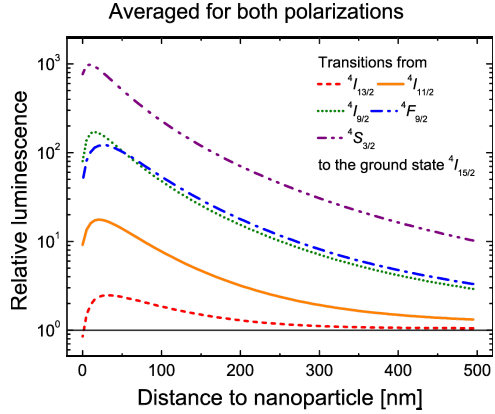

Fig. 6. Relative luminescence averaged over spherical shells around the metal nanoparticle at a defined distance and for the weighted average of SPOL and PPOL. The black solid line shows the case without metal nanoparticle.

The UC luminescence $L_i$ can be estimated from the excitation power and the number of absorbed photons $n$ that are required to populate the corresponding energy level [33, 34]. For low excitation power or low irradiance $I$, the relation $L_i \propto I^n$ was determined. Transitions from higher energy levels, consequently, benefit more from the optical field enhancement $\gamma_E$ than lower energy levels, because more photons $n$ need to be absorbed to populate higher energy levels.

For the single upconverter, the contribution of the $^4I_{11/2}$, $^4I_{9/2}$ and $^4F_{9/2}$ transitions to the ground state at an irradiance of 1000 Wm$^{-2}$ are 99.4%, 0.45% and 0.15% respectively [7]. For higher irradiances, the contribution of higher energy levels increases, which is consistent with the power law mentioned above. Higher energy levels were not included in these measurements, because only a marginal contribution to the overall luminescence is expected. Due to the different values of relative luminescence, the contribution of UC luminescence for the various energy levels alters. For the highest values of relative luminescence from SPOL (see Fig. 5 (b)), the contribution of the $^4I_{11/2}$, $^4I_{9/2}$ and $^4F_{9/2}$ transitions are 95.6%, 3.2% and 1.4%. Hence the emission spectrum of UC is modified by the gold nanoparticle. The $^4I_{11/2}$ emission centered at 980 nm, however, still dominates the overall UC luminescence enhancement.

*3.2.3. Average over the complete simulation volume*

If it is not possible to put the upconverter at a certain distance to the nanoparticles, the typical geometry would be a homogenously dispersed upconverter around the nanoparticle. For this case, we calculated the average value over all simulation points in the volume and both polarizations. The results are summarized in Table I. The luminescence is enhanced by over 121% for the energy level $^4F_{9/2}$, but only 17% for the dominant transition from $^4I_{11/2}$ to the ground state.

Table I: Average relative luminescence for a homogenously dispersed upconverter. The average was calculated over all simulation points for SPOL, PPOL and weighted for SPOL and PPOL.

| Energy level | Relative luminescence | | |
|---|---|---|---|
| (transition wavelength) | PPOL | SPOL | Average |
| $^4I_{13/2}$ (1523 nm) | 0.99 | 1.09 | 1.03 |
| $^4I_{11/2}$ (980 nm) | 1.01 | 1.49 | 1.17 |
| $^4I_{9/2}$ (805 nm) | 1.05 | 3.89 | 2.00 |
| $^4F_{9/2}$ (655 nm) | 1.10 | 4.41 | 2.21 |
| $^4S_{3/2}$ (540 nm) | 2.25 | 12.84 | 5.78 |

The results show that only at certain positions and mainly for SPOL a significant enhanced UC is feasible. However, for optimized MNP with resonances optimized with respect to $\gamma_E$ and $\gamma_{rad}$, we expect much higher improvements of the UC performance. The focus of our future work will be on investigations of how the strong increase of the emission from energy level $^4F_{9/2}$ can be transferred to the $^4I_{11/2}$ by changing the parameters of the metal nanoparticle, like shape and size, to tune its resonance into more favorable energy ranges.

## 4. Conclusion

We have presented a rate equation model to describe the UC dynamics of $\beta$-NaEr$_{0.2}$Y$_{0.8}$F$_4$. The model considers stimulated absorption and emission, spontaneous emission, multi-phonon relaxation and energy transfer. We found a good agreement between the simulations of the UC quantum yield with results from calibrated photoluminescence measurements. The UC model was coupled with simulations of a spherical gold nanoparticle. We considered changes in the optical field as well as changes of the transition probabilities, due to the coupling between the plasmon resonances of the metal nanoparticle and the Er$^{3+}$ dipole emitter. Parallel (PPOL) as well as perpendicular (SPOL) orientations of the oscillating dipole emitter to the surface of the metal nanoparticle were calculated. We investigated the influence of a spherical gold nanoparticle with a diameter of 200 nm on the UC performance in a cubic simulation volume with an edge length of 6 times the diameter of the gold nanoparticle. The relative changes of the optical field and the transition probabilities were calculated and the rate equations of the UC model were solved, with the changed parameters, for every position in the simulation volume.

The effect of the plasmon resonance on the UC quantum yield varies strongly with the location and the polarization of the oscillating dipole. The effect of the gold nanoparticle on the transition from $^4I_{11/2}$ to the ground state $^4I_{15/2}$, at a wavelength of 980 nm, is a good indicator of an enhanced UC quantum yield due to plasmon resonances, because more than 95% of the total UC quantum yield is contributed by this transition. At certain positions and for SPOL, the luminescence from the $^4I_{11/2}$ level can be enhanced by a factor of 500. For a weighted average of SPOL and PPOL, the luminescence of the transition from $^4I_{11/2}$ is increased by 15 at distances between 10 nm and 40 nm. This configuration could be achieved by a coating of the nanoparticles. With such a design it should be possible to use this enhancement to increase the efficiency of UC solar cell devices. With optimized MNP, we expect much greater improvements to the UC quantum yield. The change in the transition probabilities due to the modification of the density of states has a stronger effect on the UC performance than the optical field enhancement, because of the large difference between the two polarizations of the dipole emitter and the enhanced relative luminescence far away from the metal nanoparticle, where the optical field enhancement is only marginal.

**Acknowledgement**

The research leading to these results has received funding from the German Federal Ministry of Education and Research in the project "InfraVolt – Infrarot-Optische Nanostrukturen für die Photovoltaik" (BMBF, project numbers 03SF0401B and 03SF0401E), and from the European Community's Seventh Framework Programme (FP7/2007-2013) under grant agreement n° [246200]. S. Fischer gratefully acknowledges the scholarship support from the Deutsche Bundesstiftung Umwelt DBU.

**References**

1. T. Trupke, M. A. Green, and P. Würfel, "Improving solar cell efficiencies by up-conversion of sub-band-gap light," J. Appl. Phys. **92**, 4117-4122 (2002).
2. P. Gibart, F. Auzel, J. C. Guillaume, and K. Zahraman, "Below band-gap IR response of substrate-free GaAs solar cells using two- photon up-conversion," Japanese Journal of Applied Physics **35**, 4401-4402 (1996).
3. W. Shockley and H. J. Queisser, "Detailed balance limit of efficiency of p-n junction solar cells," J. Appl. Phys. **32**, 510-519 (1961).
4. T. Trupke, A. Shalav, B. S. Richards, P. Würfel, and M. A. Green, "Efficiency enhancement of solar cells by luminescent up-conversion of sunlight," Sol. Ener. Mater. Sol. Cells **90**, 3327-3338 (2006).
5. K. W. Krämer, D. Biner, G. Frei, H. U. Güdel, M. P. Hehlen, and S. R. Lüthi, "Hexagonal sodium yttrium fluoride based green and blue emitting upconversion phosphors," Chem. Mater. **16**, 1244-1251 (2004).
6. B. S. Richards and A. Shalav, "Enhancing the near-infrared spectral response of silicon optoelectronic devices via up-conversion," IEEE Trans. Electron Devices **54**, 2679-2684 (2007).
7. S. Fischer, J. C. Goldschmidt, P. Loeper, G. H. Bauer, R. Brueggemann, K. Kraemer, D. Biner, M. Hermle, and S. W. Glunz "Enhancement of silicon solar cell efficiency by upconversion: Optical and electrical characterization," J. Appl. Phys. **108**, 044912 (2010).
8. A. Shalav, B. S. Richards, and M. A. Green, "Luminescent layers for enhanced silicon solar cell performance: up-conversion," Sol. Ener. Mater. Sol. Cells **91**, 829-842 (2007).
9. C. Strümpel, *Application of erbium-doped up-converters to silicon solar cells* (Hartung-Gorre Verlag Konstanz, 2008), p. 144.
10. J. C. Goldschmidt, S. Fischer, P. Löper, K. W. Krämer, D. Biner, M. Hermle, and S. W. Glunz, "Experimental analysis of upconversion with both coherent monochromatic irradiation and broad spectrum illumination," Sol. Ener. Mater. Sol. Cells **95**, 1960-1963 (2011).
11. O. L. Malta, P. A. Santa-Cruz, G. F. de Sa´, and F. Auzel, "Up-conversion yield in glass ceramics containing silver," J. Solid State Chem. **68**, 314-319 (1987).
12. P. Johansson, H. Xu, and M. Käll, "Surface-enhanced Raman scattering and fluorescence near metal nanoparticles," Physical Review B **72**, 035427 (2005).
13. H. Mertens and A. Polman, "Plasmon-enhanced erbium luminescence," Appl. Phys. Lett. **89**, 1-3 (2006).
14. F. Hallermann, C. Rockstuhl, S. Fahr, G. Seifert, S. Wackerow, H. Graener, G. von Plessen, and F. Lederer, "On the use of localized plasmon polaritons in solar cells," Phys. Status Solidi A **205**, 2844-2861 (2008).
15. J. Gersten and A. Nitzan, "Spectroscopic properties of molecules interacting with small dielectric particles," J. Chem. Phys. **75**, 1139-1152 (1981).
16. F. Kaminski, V. Sandoghdar, and M. Agio, "Finite-Difference Time-Domain Modeling of Decay Rates in the Near Field of Metal Nanostructures," Journal of Computational and Theoretical Nanoscience **4**, 635-643 (2007).

17. G. Sun, J. B. Khurgin, and R. A. Soref, "Practicable enhancement of spontaneous emission using surface plasmons," Appl. Phys. Lett. **90**, 111107-111103 (2007).
18. O. L. Muskens, V. Giannini, J. A. Sánchez-Gil, and J. Gómez Rivas, "Strong Enhancement of the Radiative Decay Rate of Emitters by Single Plasmonic Nanoantennas," Nano Letters **7**, 2871-2875 (2007).
19. J. T. v. Wijngaarden and et al., "Enhancement of the decay rate by plasmon coupling for Eu 3+ in an Au nanoparticle model system," EPL (Europhysics Letters) **93**, 57005 (2011).
20. H. Zhang, Y. Li, I. A. Ivanov, Y. Qu, Y. Huang, and X. Duan, "Plasmonic Modulation of the Upconversion Fluorescence in NaYF4:Yb/Tm Hexaplate Nanocrystals Using Gold Nanoparticles or Nanoshells," Angew. Chem. Int. Ed. **49**, 2865-2868 (2010).
21. N. Liu, W. Qin, G. Qin, T. Jiang, and D. Zhao, "Highly plasmon-enhanced upconversion emissions from Au@[small beta]-NaYF4:Yb,Tm hybrid nanostructures," Chemical Communications **47**, 7671-7673 (2011).
22. S. Schietinger, T. Aichele, H.-Q. Wang, T. Nann, and O. Benson, "Plasmon-Enhanced Upconversion in Single NaYF4:Yb3+/Er3+ Codoped Nanocrystals," Nano Letters **10**, 134-138 (2009).
23. S. Fischer, H. Steinkemper, P. Löper, M. Hermle, and J. C. Goldschmidt, "Modeling upconversion of erbium doped microcrystals based on experimentally determined Einstein coeffiecienct," J. Appl. Phys. **111**, 013109 (2012).
24. F. Auzel, "Upconversion and anti-stokes processes with f and d ions in solids," Chemical Review **104**, 139-173 (2004).
25. A. Einstein, "Zur Quantentheorie der Strahlung," Physikalische Zeitschrift **18**, 121-128 (1917).
26. P. Kubelka and F. Munk, "Ein Beitrag zur Optik der Farbanstriche," Zeitschrift für technische Physik **11 a**, 593-601 (1931).
27. L. Yang and B. Kruse, "Revised Kubelka-Munk theory. I. Theory and application," Journal of the Optical Society of America A **21**, 1933-1941 (2004).
28. B. R. Judd, "Optical absorption intensities of rare-earth ions," Physical Review **127**, 750-761 (1962).
29. G. S. Ofelt, "Intensities of crystal spectra of rare-earth ions," J. Chem. Phys. **37**, 511-520 (1962).
30. J. C. Goldschmidt, *Novel Solar Cell Concepts* (Verlag Dr. Hut, München, 2009), p. 273.
31. Y. S. Kim, P. T. Leung, and T. F. George, "Classical decay rates for molecules in the presence of a spherical surface: A complete treatment," Surf. Sci. **195**, 1-14 (1988).
32. H. Mertens, A. F. Koenderink, and A. Polman, "Plasmon-enhanced luminescence near noble-metal nanospheres: Comparison of exact theory and an improved Gersten and Nitzan model," Physical Review B **76**, 115123 (2007).
33. M. Pollnau, D. R. Gamelin, S. R. Lüthi, H. U. Güdel, and M. P. Hehlen, "Power dependence of upconversion luminescence in lanthanide and transition-metal-ion systems," Physical Review B **61**, 3337-3346 (2000).
34. J. F. Suyver, A. Aebischer, S. Garcia-Revilla, P. Gerner, and H. U. Güdel, "Anomalous power dependence of sensitized upconversion luminescence," Physical Review B **71**, 125123-125121-125129 (2005).